\documentstyle[12pt]{article}
\makeatletter
\newcommand{\nn}{\nonumber}
\newcommand{\bd}{\begin{document}}
\newcommand{\ed}{\end{document}}
\newcommand{\bc}{\begin{center}}
\newcommand{\ec}{\end{center}}
\newcommand{\be}{\begin{eqnarray}}
\newcommand{\ee}{\end{eqnarray}}
\newcommand{\eqn}{\global\def\theequation}
\newcommand{\sw}{sin^2 \theta_W}
\newcommand{\fbd}{f_B}
\renewcommand{\thefootnote}{\alph{footnote}}
\def\figcap{\section*{Figure Captions\markboth
     {FIGURECAPTIONS}{FIGURECAPTIONS}}\list
     {Figure \arabic{enumi}:\hfill}{\settowidth\labelwidth{Figure 999:}
     \leftmargin\labelwidth
     \advance\leftmargin\labelsep\usecounter{enumi}}}
\let\endfigcap\endlist \relax
\def\reflist{\section*{References\markboth
     {REFLIST}{REFLIST}}\list
     {[\arabic{enumi}]\hfill}{\settowidth\labelwidth{[999]}
     \leftmargin\labelwidth
     \advance\leftmargin\labelsep\usecounter{enumi}}}
\let\endreflist\endlist \relax
\renewcommand{\thefootnote}{\alph{footnote}}
\begin{document}
\tolerance=10000
\begin{titlepage}
\begin{flushright}
{\normalsize     NHCU-HEP-94-20
\\ hep-ph/9506295}
\end{flushright}
 \null
 \vskip 1.5in
\begin{center}
 \vspace{.15in}
{\Large {\bf Axion-photon Couplings in Invisible Axion Models }}\\
  \par
 \vskip 2.5em
 {\large
  \begin{tabular}[t]{c}
        {\bf S.~L.~Cheng$^a$, C.~Q.~Geng$^b$ and W.-T.~Ni$^b$}
\\
\\
   {\em   $^a$Department of Physics, National Chung Hsing University} \\
  {\em  Taichung, Taiwan, Republic of China} \\
and\\
   {\em   $^b$Department of Physics, National Tsing Hua University} \\
  {\em  Hsinchu, Taiwan, Republic of China}
   \end{tabular}}
 \par \vskip 5.0em
 {\large\bf Abstract}
\end{center}
\setlength{\baselineskip}{5ex}

We reexamine the axion-photon couplings in various
invisible axion models motivated by the recent proposal of
using optical interferometry at the ASST facility in the SSCL
to search for axion.
We illustrate that the assignment of
$U(1)_{PQ}$ charges for the fermion fields plays an important
role in determining the couplings.
Several simple non-minimal invisible axion models with suppressed
and enhanced axion-photon couplings are constructed, respectively.
We also discuss the implications of possible new experiments to
detect solar axions by conversion to $X$-rays in a static magnetic
apparatus tracking the sun.

\end{titlepage}
\newpage

\setlength{\baselineskip}{5ex}
\pagestyle{plain}
\pagenumbering{arabic}
\setcounter{page}{2}

\section{Introduction }

$\ \ \ $
A interesting problem in the standard model is the strong
CP problem \cite{review} of
why the parameter $\theta$ which involves a $P$ and $T$-odd
term in the QCD Lagrangian, is so small: $\theta< 10^{-9}$.
A natural and elegant solution to this problem is the
Peccei-Quinn (PQ) mechanism \cite{PQ} which yields $\theta=0$ dynamically.
The spontaneous breakdown of chiral global $U(1)_{PQ}$ symmetry
gives rise to a pseudo-Goldstone boson called axion \cite{axion}.
To elude current experimental detections
it has to be very light and weakly coupled
and therefore is dubbed the ``invisible'' axion.
Astrophysical and cosmological considerations
require its mass $m_a$ to lie between $10^{-3}$ and $10^{-6}\ eV$
\cite{review}.

Recent observation of the quadrupole anisotropy in the cosmic microwave
background radiation \cite{x,x1}, favors cold dark matter
cosmologies \cite{x2}. Axions and/or other pseudo-Goldstone
bosons together with light mass neutrinos are potentially good candidates
for cold dark matter. Interests in axion search have been enhanced
recently.

There are existing many interesting ongoing and underway experiments
to search for the axion \cite{review,n7},
especially the recent proposal \cite{Ni} of using
optical interferometry at the Accelerator System String Test (ASST) facility
in the SSCL to study the velocity of light in a strong magnetic field.
Since the axion couples to photons through the anomaly,
the magnitude of the light retardation in a
magnetic field depends on the axion mass
and its coupling strength, which are model-dependent.
Measuring a deviation from the QED prediction would test
and distinguish various invisible axion models.
The sensitivity of the experiments is related to the accuracy
of the measurement of the vacuum birefringence effect. For the
accuracy of 0.3\% expected from the ultra-high sensitive interferometer
the search for axions can reach $10^{11}\ GeV$ for the inverse of
the axion-photon coupling, which is several orders of
magnitude higher than that reached by previous experiments \cite{Ni}.
More specifically, the proposed experiment aims at using SSCL ASST $170\ m$
magnet string facility with a dipole field of $6.6\ T$.
Using a Fabry-Perot cavity of laser $(\lambda =1.06\ \mu m)$
of the output power $P=50\ W$, the sensitivity limited by the shot noise
is $2.64\times 10^{-20}\ m/\sqrt{Hz}$ for
the optical pathlength variation ($\delta l$) due to different polarization
and  $1.70\times 10^{-22}/\sqrt{Hz}$ for
the variation of refraction index ($\delta n$) due to different polarization
with active length $155\ m$. For an integration time of about 30 hours,
$\Delta f=10\ \mu Hz$, it is possible to measure $\delta n$ to
$5.38\times 10^{-25}$. In a field of $6.6\ T$ magnetic field,
$\delta n$ is $1.75\times 10^{-22}$. Hence, for an integration of
30 hours, the QED birefringence effect would be detected to
0.3\%.
With this sensitivity, the mass scale $M$ that can be probed for
axions approaches
$10^{11}\ GeV$.
Longer integration will give better sensitivity.

The most recent search for solar axion \cite{n9} is to exploit their
conversion to $X$-rays in a static magnetic field for detection.
They used a magnet of length $1.8\ m$ and field strength $B=2.2\ T$.
The diameter of the axion converter is about $0.15\ m$. Data were taken for
about two hours centered around sunset of two days. The effective time
window is about $15\ mins$ per day. They set a limit of
$M>2.79\times 10^{8}\ GeV$ ($99.7\%\ C.L.$) for
$m_a<0.03\ eV$.
With a longer magnet ($5\ m$) and longer effective diameter $(3\ m)$
at $7\ T$ tracking the sun,
the sensitivity for detecting axions
can be improved by two orders of magnitude or more in $M$.
Building such a magnet is feasible.
As a matter of fact, among the two superconducting electromagnetic
thrusters for the superconducting electromagnetic propulsion ship
{\em YAMATO1}
(experimental ship),\footnote{We thank M. Wake
for providing the information of the
{\em YAMATO1}
superconducting magnets to us.}
each has six coils arranged in a
circle. The coils have an inside diameter of $0.36\ m$
and effective length of $3\ m$ with a magnetic field $3-4\ T$.
The total volume for the magnetic field of each thruster is $1.8\ m^3$.
Therefore,
the mass scale $M$ that can be probed for solar axion conversion
approaches $10^{11}\ GeV$. Larger volume and higher field or
better detector sensitivity and longer integration will give better
sensitivity. This method is comparable to the optical interferometry
method in sensitivity to $M$.
Recently, an experimental effort has been going on in Novosibirsk by using
a gimballed magnet to track the Sun with a sensitivity goal approaching
$10^{10}\ GeV$ \cite{Novosibirsk}.

In this paper, we will systematically
reexamine the axion-photon couplings in axion models.
In particular, we will explore the possibility
of having small or large couplings
in simple extensions of the original invisible axion models.

\section{The models}

$\ \ \ $
There are three minimal types of
invisible axion models: (1) Dine-Fischler-Srednicki-Zhitiniskii (DFSZ)
type \cite{DFSZ}, which has two doublets
$\phi_i\ (i=1,2)$ and one singlet $\chi$ Higgs fields;
(2) Kim-Shifman-Vainshtein-Zakharov (KSVZ) type \cite{KSVZ},
which contains one $\phi$ and one $\chi$ Higgs fields plus
a superheavy exotic $SU(2)\times U(1)$ singlet quark $Q$; and
(3) the variant invisible axion (VIA) models \cite{VIA}.
The latter modifies the variant weak scale axion models \cite{vaxion}
by adding a single Higgs field as the DFSZ models.
There are many variations of these minimal
axion models depending on the choices of the PQ charges.

Without loss of generality, we assign the following $U(1)_{PQ}$ transformations
for the Higgs and the fermion fields:
\be
&&
\phi_1\to \exp (-ih_1)\phi_1\,,\
\phi_2\to \exp (ih_2)\phi_2\,,\
\nn\\
&&
\chi\to \exp (-i(h_1+h_2)/2)\chi\,,\
\nn\\
&&
d_R^j \to \exp (iX_d^j)d_R^j\,,\
u_R^j \to \exp (iX_u^j)u_R^j\,,\
e_R^j \to \exp (iX_e^j)e_R^j\,,\
\ee
in the DFSZ and VIA models with $j=1,\cdots, N_g$ being the generation
index, and
\be
&&
\chi\to \exp (iQ_{PQ})\chi\,,\
\nn\\
&&
Q_L \to \exp (iQ_{PQ}/2)Q_L\,,\
Q_R \to \exp (-iQ_{PQ}/2)Q_R\,,\
\ee
in the KSVZ models.
The PQ transformations in the DFSZ models
are generation blind and they distinguish only the up- and down-type
quarks and charged leptons. Thus, in these models, we have
\be
&&X_u\:\equiv X_u^i\,,\
X_d\:\equiv X_d^i\,,\
X_e\:\equiv X_e^i\,,\
\nn\\
&&X_u\:\neq\: -X_d
\ee
with $i=1,\cdots,N_g$. The down-type (up-type) quarks can get masses from the
Yukawa interactions by choosing $h_{1(2)}=X_{d(u)}$.
However, there is a freedom for the lepton Yukawa couplings, depending
on the PQ charge of $X_e$. This freedom leads to two minimal DFSZ models:
\be
{\rm DFSZ\ I}&:& \ X_e\:=\:X_d\,,\ \ \ \ \ \ \ \ \ \ \ \ \ \ \ \ \ \ \
\nn\\
{\rm DFSZ\ II}&:& \ X_e\:=\:-X_u\,.
\ee
The first model is the original DFSZ model.
We now consider the cases that the charged lepton masses arise from
a third Higgs doublet, $\phi_3$, to take into account the lightness
of their masses comparing with the quark ones.
We assign the PQ charge of $\phi_3$ as
\be
h_3 &=&-X_e\:\,\neq\:\, |X_u|\,,\ |X_d|\,.\ \ \ \ \ \ \ \ \ \ \ \ \ \ \
\ee
We refer the assignment in Eq. (5) as DFSZ III.
For the VIA models, $U(1)_{PQ}$ is no longer
family blind.
As an example, we take $N_g=3$ and the PQ charges in Eq. (1) as
\be
(X_u^1\,,\ X_u^2\,,\ X_u^3)&=& (-X_d\,,\ -X_d\,,\ X_u)\,,
\nn\\
(X_d^1\,,\ X_d^2\,,\ X_d^3)&=& (X_d\,,\ X_d\,,\ X_d)\,,
\nn\\
(X_e^1\,,\ X_e^2\,,\ X_e^3)&=& (X_d\,,\ X_d\,,\ X_d)\,.
\ee
In this VIA model, the fact that the top quark is much heavier
than other flavors could be understood if $v_1\ll v_2$, where $v_{1(2)}$
is VEV of $\phi_{1(2)}$.
Finally we recall the composite axion model suggested by Kim \cite{Kim}.
In this model
the transformations of the exotic Dirac fermions,
under the group $SU(N)_{MC}\times SU(3)_C\times U(1)_Y\times U(1)_{PQ}$ are
chosen to be
\be
& (N,3,a;1)+(N,1,b;-3)\,;& \nn\\
& (\bar{N},\bar{3},-a;1)+(\bar{N},1,-b;-3)\,,&
\ee
where $N$ is the dimension of the metacolor representation.
Clearly, $U(1)_{PQ}$ is indeed a PQ symmetry since it has a color anomaly.

\section{The axion-photon interactions}

$\ \ \ $
We write the effective axion-photon Lagrangian as
\be
{\cal L}_{a\gamma\gamma}
&=& {1\over 4}g_{a\gamma\gamma} aF_{\mu\nu}\tilde{F}^{\mu\nu}
\ee
where $F_{\mu\nu}$ is the electromagnetic field tensor and
$\tilde{F}^{\mu\nu}$ its dual, $a$  the axion field, and
$g_{a\gamma\gamma}$  the coupling constant.
For a static magnetic field and real photon, the interaction becomes
\be
{\cal L}_{a\gamma\gamma}
&=& {1\over M} a\;\vec{E}\cdot \vec{B}_{ext}
\ee
where $\vec{E}$ and $\vec{B}_{ext}$ are the electric and magnetic fields
and $M$ is the energy scale defined as the inverse coupling constant,
\be
M &\equiv & {1\over g_{a\gamma\gamma}},
\ee
which has dimensions of energy. Clearly, the axion only interacts
with photon wave component that is parallel to an external magnetic
field.
One finds \cite{Kaplan}
\be
g_{a\gamma\gamma} &=& {\alpha m_a\over 2\pi f_{\pi}m_{\pi}}
{1+Z\over \sqrt{Z}}
\left[
{A_{PQ}^{em}\over A_{PQ}^C}-{2\over 3}{(4+Z)\over (1+Z)}
\right]
\ee
where $Z\equiv m_u/m_d$ and
$A_{PQ}^{em}$ and $A_{PQ}^C$ are the anomaly factors
given by
\be
A_{PQ}^{em}&=& TrQ_{em}^2Q_{PQ}
\nn\\
\delta_{ab}A_{PQ}^C&=&Tr\lambda_a\lambda_bQ_{PQ}
\ee
with $\lambda_a$ being the color generators and
$Tr\lambda_a\lambda_b={1\over 2}\delta_{ab}$.
Using $Z\simeq 0.56$, $f_{\pi}=93\ MeV$,
and $m_{\pi}=134\ MeV$, we get
\be
g_{a\gamma\gamma}
&=& 1.36\times 10^{-11}{\alpha\over m_e}\left({m_a\over 1\;eV}\right)
\left[ {A_{PQ}^{em}\over A_{PQ}^C}-1.95\right]\,.
\ee
We note that the ratio $R\equiv A_{PQ}^{em}/A_{PQ}^C$ is model-dependent.
The coupling $g_{a\gamma\gamma}$ becomes minimal when
$R$ is close to $1.95$ and, on the other hand,
it can be large for the ratio $R$ being negative or large.

We now examine the axion-photon coupling in the various invisible axion
models given in the previous section. We are particularly interested in
finding the cases that could result in small or large
$g_{a\gamma\gamma}$.
 From Eq. (1), we find for the DFSZ and VIA models
\be
A_{PQ}^{em}&=& \sum_j\left({4\over 3}X_u^j+{1\over 3}X_d^j+X_e^j\right)\,,
\nn\\
A_{PQ}^C&=&
{1\over 2}\sum_j(X_u^j+X_d^j)\,.
\ee
This leads to
\be
R_{DFSZ}  &=& {2\over 3}
(4X_u+X_d+3X_e)/(X_u+X_d)
\ee
for the DFSZ models and
\be
R_{VIA} &=& 8/3
\ee
for the VIA model in Eq. (6).
For the DFSZ I and II and VIA models,
the values of $R$ are fixed and the couplings are given by
\be
g_{a\gamma\gamma}&=& (1.4\,,\ -2.6\,,\ 1.4)\times 10^{-10}
\left({m_a\over 1eV}\right)\:GeV^{-1}\,,
\ee
respectively.
However, $R$ in the DFSZ III varies as we change  $X_e$ assignment.
It is easily seen that
a model with $R=2$ can be constructed if we chose
\be
X_e\:=\:1/3\,,\ \:X_u\:=\:X_d\:=\:1\,,
\ee
in which we get
\be
g_{a\gamma\gamma}&=& 9.9\times 10^{-12}\left({m_a\over 1eV}\right)\:GeV^{-1}\,,
\ee
from Eqs. (13) and (15).
We can also have models with large $g_{a\gamma\gamma}$. For example,
if we set
\be
X_e\:=\:-1\,,\ \:X_u\:=\:-1/3\,,\ \:X_d\:=\:2/3\,,
\ee
corresponding to $R=-22/3$, we get
\be
g_{a\gamma\gamma}&=& -1.9\times 10^{-9}\left({m_a\over 1eV}\right)\:GeV^{-1}\,.
\ee
The absolute value of $g_{a\gamma\gamma}$
in Eq. (21) is about a factor
of 200 larger than that in Eq. (19).
The inverse axion-photon coupling strength $M\equiv g_{a\gamma\gamma}^{-1}$
in various cases are shown in Table 1.

Similarly, we have \cite{Kaplan}
\be
R &=& 6 Q_{em}^2\ \ {\rm and}\ \ 6(a^2-b^2)\,
\ee
for the KSVZ and composite axion models, respectively. Here
$Q_{em}$ is the electric charge of the heavy quark in the KSVZ models.
As emphasized by Kaplan in Ref. \cite{Kaplan}, due to the arbitrariness
of the charges in Eq. (22) it is possible
to create $R=2$ in the composite axion model naturally.
As completeness
we list it as well as other cases in Table 1.

As we can see from Table 1, it is impossible to detect
the axion in the models with suppressed photon-axion couplings
even with the new proposal of using optical interferometry at SSCL
ASST facility as well as the improved experiment of $X$-ray conversion of
solar axions. However, it could be accessible for
the models which have enhanced photon couplings when further improvement
on sensitivity is made.

\section{Conclusions}

$\ \ \ $
In the light of the recent proposal of using interferometry at
SSCL ASST facility to search for axions and pseudoscalar particles \cite{Ni},
we have systematically studied the axion-photon couplings in various
invisible axion models.
We have demonstrated that $g_{a\gamma\gamma}$  depends strongly on
the assignment of $U(1)_{PQ}$ charges, and
found that simple extensions of the minimal DFSZ models
could lead to suppressed as well as enhanced axion-photon couplings.

Although the new proposal \cite{Ni} aims at five orders of magnitude
improvement on the sensitivity of the energy scale $M$, it is still
not enough to detect the axions even in the models with
enhanced
photon-axion couplings. It may be marginal for the searches when
further improvements on the sensitivity are made.
The situation is similar for the method of $X$-ray conversion of
solar axions.
If the experiment finds deviations from the Quantum Electrodynamics at
the initial stage, the effects must be new and not from the
invisible axion models considered.
Finally, we remark that
with five orders of magnitude improvement, the regions of interest are rather
wide. Critical examinations in other  pseudo-scalar/scalar models as well as
those with potential weak two-photon couplings would be mostly interesting.

\vskip .25cm
\noindent
{\bf Acknowledgments}

This work was supported in part
by the National Science Council of Republic of China
under Grants NSC-83-0208-M-007-118 (C.Q.G) and
NSC-83-0208-M-007-126 (W.-T.N).

\newpage

\begin{table}
\caption{The inverse axion-photon coupling strength $M$
in invisible axion models}
\vspace{.5cm}
\renewcommand{\arraystretch}{2}
\noindent\begin{tabular*}{15.cm}{@{\extracolsep{\fill}}cccccccccr}
\hline \hline
& Models  &&
$\left({m_a\;eV\over 10^{-5}}\right) \;\left( {M\ GeV \over 10^{14}}\right)$
&& Remark & \\
\hline
& DFSZ I && $7.1$ && $X_e=X_d$ &\\
& DFSZ II && $3.9$ && $X_e=-X_u$ &\\
& DFSZ III && $17$ && $X_e=0\,,\ X_u=X_d$ &\\
& && $101$ && $X_e=1/3\,,\ X_u=X_d=1$ &\\
& && $0.54$ && $X_e=-1\,,\ X_u=-1/3\,,\ X_d=2/3$ &\\
\hline
& VIA && $7.1$ && &\\
\hline
& KSVZ && $7.1$ && $Q_{em}=2/3$ &\\
& && $3.9$ && $Q_{em}=-1/3$ &\\
& && $1.2$ && $Q_{em}=1$ &\\
& && $2.6$ && $Q_{em}=0$ &\\
\hline
&Composite && $0.63$ && $a=0\,,\ b=1$ &\\
&&& $101$ && $a=2/3\,,\ b=\pm1/3$ &\\
\hline \hline
\vspace{.05cm}
\end{tabular*}
$^*$ We take the absolute value of $M$.
\end{table}

\ed
\newpage

\begin{thebibliography}{99}
\addcontentsline{toc}{chapter}{Bibliography}

\bibitem{review}
For reviews, see J.E. Kim, {\it Phys. Rep.} {\bf 150} (1987) 1; H.Y. Cheng,
{\it Phys. Rep.} {\bf 158} (1988) 1; R.D. Peccei, in {\sl CP Violation}, ed.
by C. Jarlskog, (World Scientific, Singapore, 1989), p. 503;
M.S. Turner, {\it Phys. Rep.} {\bf 197} (1990) 67;
G.G. Rafelt, {\it Phys. Rep.} {\bf 198} (1990) 1; and references therein.

\bibitem{PQ} R.D. Peccei and H.R. Quinn, {\it Phys. Rev. Lett.}
{\bf 38} (1977) 1440; {\it Phys. Rev. D} {\bf 16} (1977) 1791.
\bibitem{axion} S. Weinberg, {\it Phys. Rev. Lett.} {\bf 40} (1978) 223;
F. Wilczek, {\it Phys. Rev. Lett.} {\bf 40} (1978) 279.

\bibitem{x} G. F. Smoot {\it et al.},{\it  Astrophys. J.} {\bf 396}
(1992) L1.

\bibitem{x1} S. Hancock {\it et al.}, {\it Nature} (London) {\bf 367}
(1994) 333.

\bibitem{x2} R. L. Davis, {\it Phys. Rev. Lett.} {\bf 69} (1992) 1856.

\bibitem{n7} G. Cantatore {\it et al.}, {\it Nucl. Phys.} B (Proc. Suppl.)
{\bf 35} (1994) 180.

\bibitem{Ni}
Cf. {\sl ``Light Retardation in a High Magnetic Field and
Search for Light Scalar/Pseudo-scalar Particles Using
Ultra-sensitive Interferometry''},
{\it Joint Proposal} Submitted to the National Science Council
of the Republic of China and Department of Energy of the
United States of America (April, 1994); see also
W.-T. Ni {\it et al.}, ``{\sl Test of Quantum Electrodynamics
and Search for Light Scalar/Pseudoscalar Particles Using
Ultra-high Sensitive Interferometers}'', NHCU-HEP-94-12,
in Proceedings of the Meeting on General Relativity, July 23-30, 1994,
Stanford, CA (World Scientific, Singapore, 1995) (in press).

\bibitem{n9} D. M. Lazarus {\it et al.}, {\it Phys. Rev. Lett.} {\bf 69}
(1992) 2333.

\bibitem{Novosibirsk} P.V. Vorobyov and I.V. Kolokolov,
``{\sl Detectors for the Cosmic Axionic Wind}'', ASTRO-PH-9501042 (1995).

\bibitem{DFSZ} M. Dine, W. Fischler and M. Srednicki,
{\it Phys. Lett.} {\bf B104} (1981) 199;
A.P. Zhitnitskii, {\it Sov. J. Nucl. Phys.} {\bf 31} (1980) 260.
\bibitem{KSVZ} J.E. Kim, {\it Phys. Rev. Lett.} {\bf 43} (1979) 103;
M.A. Shifman, A.I. Vainshtein, and V.I. Zakharov,
{\it Nucl. Phys.} {\bf B166} (1980) 493.
\bibitem{VIA} C. Q. Geng and J. N. Ng, {\it Phys. Rev.} {\bf D39} (1989) 1449.
\bibitem{vaxion} R.D. Peccei, T.T. Wu and T. Yanagida,
{\it Phys. Lett.} {\bf B172}
(1986) 435; L.M. Krauss and F. Wilczek, {\it Phys. Lett.} {\bf B173}
(1986) 189.
\bibitem{Kim} J.E. Kim, {\it Phys. Rev.} {\bf D31} (1985) 1733.
\bibitem{Kaplan}
W.A. Bardeen and S.-H.H. Tye, {\it Phys. Lett.} {\bf B74} (1978) 229;
D.D. Kaplan, {\it Nucl. Phys.} {\bf B260} (1985) 215;
M. Srednicki, {\it Nucl. Phys.} {\bf B260} (1985) 689.

\end{thebibliography}
